\documentclass[aps,prd,twocolumn,showpacs,showkeys,amsmath,amssymb]{revtex4}
\usepackage{graphicx}
\usepackage{dcolumn}
\usepackage{bm}

\begin{document}

\title{Diquarks, Pentaquarks and Dibaryons}
\author{Shi-Lin Zhu}
\email{zhusl@th.phy.pku.edu.cn}
\affiliation{%
Department of Physics, Peking University, BEIJING 100871, CHINA}

\date{\today}

\begin{abstract}

We explore the connection between pentaquarks and dibaryons
composed of three diquarks in the framework of the diquark model.
With the available experimental data on H dibaryon, we estimate
the Pauli blocking and annihilation effects and constrain the
$P=-$ pentaquark $SU(3)_F$ singlet mass. Using the $\Theta^+$
pentaquark mass, we estimate $P=-$ dibaryon mass.

\end{abstract}


\maketitle

\pagenumbering{arabic}

\section{Introduction}\label{sec0}

Baryons in the conventional quark model are color singlets
composed of three quarks. So their color wave function is
anti-symmetric. Pauli principle requires the total wave function
of three quarks be anti-symmetric. For the $L=0$ ground state
baryons, their orbital wave function is symmetric. Therefore their
spin-flavor wave function is totally symmetric, corresponding to
the nucleon octet and Delta decuplet with positive parity. The
mass splitting between the members of the $SU(6)$ multiplet is
caused by either the color-spin interaction from the gluon
exchange or the flavor-spin interaction from the pseudoscalar
meson exchange.

Quark model has been very successful in the classification of
baryons \cite{pdg}. However, quantum chromo-dynamics (QCD) as the
underlying theory of strong interaction, allows a much richer
baryon spectrum. Especially there may exist hybrid baryons (qqqG)
and multiquark baryons such as pentaquarks (qqqq$\bar q$),
dibaryons (qqqqqq) etc. Since Jaffe proposed the H dibaryon in
1977 \cite{jaffeold}, there has been extensive experimental search
of this state. There also exist discussions of other possible
dibaryons in literature \cite{wang,ihep}. Up to now, none of these
non-conventional baryon states has been established experimentally
except pentaquarks.

The surprising discovery of $\Theta^+$ pentaquark
\cite{leps,diana} last year is one of the most important events in
hadron physics for the past decades. There have appeared more than
two hundred pentaquark papers in literature within one year. Its
quantum number, internal structure, decay mechanism and underlying
dynamics are under heated debate
\cite{diak,jaffe,zhu,positive,lipkin,lattice,g2,ellis,cohen,song,zzy,he,huang,zhang,thomas,frank}.

Jaffe and Wilczek proposed the diquark picture for pentaquarks
\cite{jaffe}. The diquark is very similar to an anti-quark in many
aspects. This feature leads to deep connection between pentaquarks
and dibaryons which are composed of three diquarks. In this short
note, we will explore this connection.

\section{$P=+$ pentaquarks {\sl vs} $P=-$ dibaryons}\label{sec3}

Within the framework of the diquark model, we discuss the
connection between $P=+$ pentaquarks and those $P=-$ dibaryons
which are composed of three diquarks and contain one orbital
excitation between diquarks.

Jaffe and Wilczek proposed that the $\Theta^+$ pentaquark is
composed of two diquarks and one strange anti-quark \cite{jaffe}.
They argued that the light quarks are strongly correlated. Two
light quarks tend to form a scalar diquark in the ${\bar 3}_c,
{\bar 3}_F$ representation whenever possible. The lighter the
quark mass, the stronger the correlation. The one-gluon-exchange
interaction and the instanton induced interaction seem to support
such an idea.

Since the pentaquark is a color singlet, the color wave function
of the two diquarks within the pentaquark must be antisymmetric
$\bf{3}_C$. In order to get an exotic anti-decuplet, the two
scalar diquarks combine into the symmetric SU(3) $\bf{\bar{6}_F}$
: $[ud]^2$, $[ud][ds]_+$, $[su]^2$, $[su][ds]_+$, $[ds]^2$, and
$[ds][ud]_+$. Bose statistics demands symmetric total wave
function of the diquark-diquark system, which leads to the
antisymmetric spatial wave function with one orbital excitation.
The resulting anti-decuplet and octet pentaquarks have
$J^P={1\over 2}^+, {3\over 2}^+$. The resulting flavor wave
functions are collected in Table \ref{tab2}.

\begin{table*}[h]
\begin{center}
\begin{tabular}{cc||cc}\hline
($Y,I,I_3$)       &$\bf{\bar{10}}$ &($Y,I,I_3$)&$\bf{8}$\\
\hline
(2,0,0)            &$[ud]^2\bar{s}$ &$-$&$-$\\
(1,$\frac{1}{2}$,$\frac{1}{2}$)&$\sqrt{\frac{2}{3}}[ud][us]_+\bar{s}+\sqrt\frac{1}{3}[ud]^2\bar{d}$
     &(1,$\frac{1}{2}$,$\frac{1}{2}$)&$\sqrt\frac{1}{3}[ud][us]_+\bar{s}-\sqrt{\frac{2}{3}}[ud]^2\bar{d}$\\
(1,$\frac{1}{2}$,-$\frac{1}{2}$)&$\sqrt{\frac{2}{3}}[ud][ds]_+\bar{s}+\sqrt\frac{1}{3}[ud]^2\bar{u}$
    &(1,$\frac{1}{2}$,$\frac{1}{2}$)&$\sqrt\frac{1}{3}[ud][ds]_+\bar{s}-\sqrt{\frac{2}{3}}[ud]^2\bar{u}$\\
(0,1,1)
&$\sqrt{\frac{2}{3}}[ud][us]_+\bar{d}+\sqrt\frac{1}{3}[us]^2\bar{s}$
     &(0,1,1)&$\sqrt{\frac{1}{3}}[ud][us]_+\bar{d}-\sqrt\frac{2}{3}[us]^2\bar{s}$\\
(0,1,0) &$\sqrt{\frac{1}{3}}([ud][ds]_+\bar{d}+[ud][us]_+\bar{u}$
     &(0,1,0)&$\sqrt{\frac{1}{6}}([ud][ds]_+\bar{d}+[ud][us]_+\bar{u})$\\
  &$+[us][ds]_+\bar{s})$&&$-\sqrt{\frac{2}{3}}[us][ds]_+\bar{s}$\\
(0,1,-1)
&$\sqrt{\frac{2}{3}}[ud][ds]_+\bar{u}+\sqrt\frac{1}{3}[ds]^2\bar{s}$
     &(0,1,-1)&$\sqrt{\frac{1}{3}}[ud][ds]_+\bar{u}-\sqrt\frac{2}{3}[ds]^2\bar{s}$\\
(-1,$\frac{3}{2}$,$\frac{3}{2}$)&$[us]^2\bar{d}$ &$-$&$-$\\
(-1,$\frac{3}{2}$,$\frac{1}{2}$)&$\sqrt{\frac{2}{3}}[us][ds]_+\bar{d}+\sqrt\frac{1}{3}[us]^2\bar{u}$
     &(-1,$\frac{1}{2}$,$\frac{1}{2}$)&$\sqrt{\frac{1}{3}}[us][ds]_+\bar{d}-\sqrt\frac{2}{3}[us]^2\bar{u}$\\
(-1,$\frac{3}{2}$,-$\frac{1}{2}$&$\sqrt{\frac{2}{3}}[ds][us]_+\bar{u}+\sqrt\frac{1}{3}[ds]^2\bar{d}$
     &(-1,$\frac{1}{2}$,-$\frac{1}{2}$)&$\sqrt{\frac{1}{3}}[ds][us]_+\bar{u}-\sqrt\frac{2}{3}[ds]^2\bar{d}$\\
(-1,$\frac{3}{2}$,-$\frac{3}{2}$)&$[ds]^2\bar{u}$ &$-$&$-$\\
$-$&$-$&(0,0,0)&$\sqrt\frac{1}{2}([ud][ds]_+\bar{d}-[ud][us]_+\bar{u})$\\
 \hline
\end{tabular}
\end{center}
\caption{Flavor wave functions in Jaffe and Wilczek's model
\cite{jaffe}.
$[q_1q_2][q_3q_4]_+=\sqrt{\frac{1}{2}}([q_1q_2][q_3q_4]+[q_3q_4][q_1q_2])$
or $[q_1q_2]^2=[q_1q_2][q_1q_2]$ is the diquark-diquark part.
}\label{tab2}
\end{table*}

Throughout our discussion we assume exact isospin symmetry. We
denote the up and strange quark mass by $m_u, m_s$ and the [ud],
[us] diquark mass by $m_{[ud]}, m_{[us]}$. Since the same quark
exists in the two diquarks, Pauli blocking effect may raise the
spectrum by $E^{L=1}_{pb}$. However the centrifugal barrier from
the orbital excitation makes two diquarks far apart. One expects
that the Pauli blocking effect is less significant for $P=+$
pentaquarks than for $P=-$ pentaquarks.

In contrast, the quark and anti-quark annihilation effect tends to
lower the spectrum. There are two kinds of possible annihilation
mechanism. For example, the $\bar u$ may annihilate with the up
quark in either [ud] or [us] diquark. Such a mechanism lowers the
pentaquark mass by $E_{ann}$. The second possibility is that the
$\bar u$ and down quark in the [ud] diquark annihilates into a
virtual K or $K^\ast$, which may also lower the pentaquark mass by
$E^\prime_{ann}$. $E_{ann}$ is probably greater than
$E^\prime_{ann}$. After taking into account of the Pauli blocking
and annihilation effects, $\Theta^+$ and $\Xi^{--}$ masses are
\begin{equation}\label{theta}
M_{\Theta^+}=2m_{[ud]}+m_s +2E^{L=1}_{pb}-4E^\prime_{ann}+E_L
\end{equation}
\begin{equation}
M_{\Xi^{--}}=2m_{[us]}+m_u +2E^{L=1}_{pb}-4E^\prime_{ann}+E_L
\end{equation}
We list $P=+$ pentaquark masses in Table \ref{tab3}.

\begin{table*}[h]
\begin{center}
\begin{tabular}{cc||cc}\hline
($Y,I,I_3$)       &$\bf{\bar{10}}$ &($Y,I,I_3$)&$\bf{8}$\\
\hline
(2,0,0)            &$2m_{[ud]}+m_s +2E^{L=1}_{pb}-4E^\prime_{ann}+E_L$ &$-$&$-$\\
(1,$\frac{1}{2}$,$\pm \frac{1}{2}$)&${1\over
3}(4m_{[ud]}+2m_{[us]}+2m_s +m_u$
&(1,$\frac{1}{2}$,$\pm\frac{1}{2}$)&${1\over
3}(5m_{[ud]}+m_{[us]}+m_s
+2m_u $\\
&$+4E^{L=1}_{pb}-4E_{ann}-8E^\prime_{ann})+E_L$ &&$+5E^{L=1}_{pb}-5E_{ann}-7E^\prime_{ann})+E_L $\\
(0,1,$\pm 1$) &${1\over 3}(2m_{[ud]}+4m_{[us]}+m_s +2m_u$
&(0,1,$\pm 1$)&${1\over 3}(m_{[ud]}+5m_{[us]}+2m_s
+m_u$\\
 &$+4E^{L=1}_{pb}-4E_{ann}-8E^\prime_{ann})+E_L$ &&$+4E^{L=1}_{pb}-5E_{ann}-7E^\prime_{ann})+E_L$\\
(0,1,0) &${1\over 3}(2m_{[ud]}+4m_{[us]}+m_s +2m_u$
     &(0,1,0)&${1\over 3}(m_{[ud]}+5m_{[us]}+2m_s
+m_u$\\
&$+3E^{L=1}_{pb}-6E_{ann}-6E^\prime_{ann})+E_L$
     &&$+3E^{L=1}_{pb}-6E_{ann}-6E^\prime_{ann})+E_L$\\
(-1,$\frac{3}{2}$,$\pm \frac{3}{2}$)&$2m_{[us]}+m_u +2E^{L=1}_{pb}-4E^\prime_{ann}+E_L$ &$-$&$-$\\
(-1,$\frac{3}{2}$,$\pm \frac{1}{2}$)&$2m_{[us]} +m_u$
     &(-1,$\frac{1}{2}$,$\pm \frac{1}{2}$)&$2m_{[us]}
+m_u$\\
&$+{1\over 3}(4E^{L=1}_{pb}-4E_{ann}-8E^\prime_{ann})+E_L$
     &&$+{1\over 3}(5E^{L=1}_{pb}-5E_{ann}-7E^\prime_{ann})+E_L$\\
$-$&$-$&(0,0,0)&$m_{[ud]}+m_{[us]}+m_u +E^{L=1}_{pb}-2E_{ann}-2E^\prime_{ann}+E_L$\\
 \hline
\end{tabular}
\end{center}
\caption{$P=+$ pentaquark masses with the correction from Pauli
blocking and the annihilation effects. $E_L$ is the orbital
excitation energy.}\label{tab3}
\end{table*}

$\Theta^+$ pentaquark is interpreted as a bound state of two
diquarks and one anti-quark by Jaffe and Wilczek \cite{jaffe}. Its
mass is as low as 1530 MeV even with one orbital excitation. One
may wonder whether one can get a low lying dibaryon with $L=1$
after replacing the anti-quark in $\Theta^+$ by a diquark.

Now let's discuss $P=-$ dibaryons composed of three scalar
diquarks with $L=1$. Its color wave function is anti-symmetric.
Its spin wave function is symmetric since diquarks are scalars.
Bose statistics requires the total wave function is symmetric.
Hence the product of the flavor and orbital wave function is
anti-symmetric. Suppose there is one orbital excitation between
two diquarks: A and B. The flavor wave function of the diquark
pair A and B must be symmetric, which is the same as in the $P=+$
pentaquarks. When the orbital wave function is mixed symmetric (or
anti-symmteric), the flavor wave function must be mixed
anti-symmteric (or symmetric). This situation is very similar to
the $L=1$ baryon multiplet in the $SU(6)_{FS}$ $70_{FS}$
representation. The only difference is that the diquark is a
scalar. Simple group theory tells us that the resulting $P=-$
dibaryons are in the $8_F$ representation.

To some extent one may correspond [ud], [us], [ds] diquarks to
$\bar S, \bar D, \bar U$ respectively. We classify the dibaryon
type depending on its $\bar S, \bar D, \bar U$ content. For
example, the quark content of the proton-type dibaryon is $\bar
U\bar U \bar D$ or $[ds][ds][us]$. We use the lower index "6" to
denote the dibaryon. For the $\Lambda$-type (or $\Sigma^0$-type)
dibaryon $\Lambda_6$ (or $\Sigma^0_6$) with the quark content
$[ud][us][ds]$, its mass can be estimated as
\begin{equation}
M_{\Lambda_6,
\Sigma^0_6}=2m_{[us]}+m_{[ud]}+E_L+2E_{pb}^{L=0}+E_{pb}^{L=1}
\end{equation}
For the $\Xi$-type dibaryon $\Xi_6$ with the quark content
$[ud][ud][us]$,
\begin{equation}
M_{\Xi_6}=2m_{[ud]}+m_{[us]}+E_L+2E_{pb}^{L=0}+2E_{pb}^{L=1}
\end{equation}
For the nucleon-type dibaryon $N_6$ with the quark content
$[us][us][ds]$,
\begin{equation}
M_{N_6}=3m_{[us]}+E_L+2E_{pb}^{L=0}+2E_{pb}^{L=1}
\end{equation}
For the $\Sigma^\pm$-type dibaryon $\Sigma^\pm_6$, its mass can be
estimated as
\begin{equation}
M_{\Sigma^\pm_6}=2m_{[us]}+m_{[ud]}+E_L+2E_{pb}^{L=0}+2E_{pb}^{L=1}
\end{equation}

\section{$P=-$ pentaquarks {\sl vs} $P=+$ dibaryons}\label{sec2}

Let's move on to those dibaryons which are composed of three
diquarks and have no orbital excitation. Three ${\bar 3}_c$
diquarks combine into a color singlet so their color wave function
is antisymmetric. Diquarks are scalars. They obey Bose statistics.
Their total wave function should be symmetric. Since there is no
orbital excitation between scalar diquarks, their spin and spatial
wave functions are symmetric. Hence their flavor wave function
must be totally anti-symmetric. I.e., the resulting dibaryon is a
$SU(3)_F$ singlet with positive parity, which is nothing but the H
dibaryon proposed by Jaffe long time ago \cite{jaffeold}. Another
P=+ dibaryon with two P-waves between the diquarks and $L=0$ could
also be low lying \cite{fec}.

Within the diquark framework, it was pointed out that lighter
pentaquarks can be formed if the two scalar diquarks are in the
antisymmetric $SU(3)_F$ $\bf 3$ representation \cite{zhang,talk}:
$[ud][su]_-$, $[ud][ds]_-$, and $[su][ds]_-$, where
$[q_1q_2][q_3q_4]_-=\sqrt{\frac{1}{2}}([q_1q_2][q_3q_4]-[q_3q_4][q_1q_2])$.
No orbital excitation is needed to ensure the symmetric total wave
function of two diquarks since the spin-flavor-color part is
symmetric. The total angular momentum of these pentaquarks is
$\frac{1}{2}$ and the parity is negative. There is no accompanying
$J={3\over 2}$ multiplet. The two diquarks combine with the
antiquark to form a $SU(3)_F$ octet and singlet pentaquark
multiplet: ${\bar 3}_F \otimes 3_F = \bf{8}_F$ $\oplus$
$\bf{1}_F$.

We want to emphasize that the above pentaquark singlet with
negative parity is very similar to the H dibaryon. Its flavor wave
function reads
\begin{equation}\label{singlet}
\frac{1}{\sqrt{3}}\left(
[ud][su]_-\bar{u}+[ds][ud]_-\bar{d}+[su][ds]_-\bar{s}\right)
\end{equation}
Since the same quark exists within two diquarks, Pauli blocking
effect may raise the spectrum by $E^{L=0}_{pb}$. In contrast, the
quark and anti-quark annihilation effect tends to lower the
spectrum by $E_{ann}$. Since there is no orbital excitation, the
diquarks are in S-wave. $E^{L=0}_{pb}$ can be quite significant
and $E^{L=0}_{pb}>> E^{L=1}_{pb}$.

The $P=-$ pentaquark singlet mass may be estimated as
\begin{eqnarray}\label{1}\nonumber
M_1 = &{1\over 3}\left( 2m_u+m_s +2m_{[ud]}+4m_{[us]}\right) \\
&+E^{L=0}_{pb}-2E_{ann}-2E^\prime_{ann}
\end{eqnarray}

Replacing the antiquark in Eq. (\ref{singlet}) by the
corresponding diquark we arrive at the H dibaryon with the diquark
content $[ud][us][ds]$. Its mass reads
\begin{equation}
M_H = m_{[ud]}+2m_{[us]}+3E^{L=0}_{pb}
\end{equation}

\section{Discussion}\label{sec4}

We follow Ref. \cite{jaffe} and use $m_u=360$ MeV, $m_s=460$ MeV,
$m_{[ud]}=420$ MeV, and $m_{[us]}=580$ MeV. If we naively ignore
the Pauli blocking and annihilation effects, we get
\begin{equation}
M_{\Lambda_6}=M_{\Theta^+}+2m_{[us]}-m_{[ud]}-m_s=1710 \mbox{MeV}
\end{equation}
where we have used $M_{\Theta^+}=1530$ MeV \cite{leps}. Such a low
lying dibaryon with negative parity is clearly in conflict with
the experimental data. In other words, the Pauli blocking and
annihilation effects are important.

We may make a rough estimate of $E^{L=0}_{pb}$ from available
experimental information on H dibaryon. If H particle really
exists, it must be a very loosely bound state which is close to
the $\Lambda\Lambda$ threshold. Its binding energy must be less
than a few MeV according to the recent doubly $\Lambda$
hyper-nuclei experiments \cite{bnl,kek}. In fact, the lower bound
of H dibaryon mass was pushed to be $M_H \ge 2224$ MeV. Then we
get
\begin{equation}\label{2}
E^{L=0}_{pb}\approx 215 \mbox{MeV}
\end{equation}
It's important to note that $E^{L=0}_{pb}$ is correlated with the
diquark mass. We may adjust the values of $m_{[ud]}, m_{[us]}$
within a reasonable range to get a smaller $E^{L=0}_{pb}$.

On the other hand, both $E_{ann}$ \cite{bicudo} and
$E^\prime_{ann}$ \cite{zzy} may be important numerically. For a
rough estimate, we use $E_{ann}\approx (50\sim 100)$ MeV,
$E^\prime_{ann}\approx (10\sim 30)$ MeV. The orbital excitation
energy $E_L$ is typically around $240$ MeV. From Eq.
(\ref{theta}), we get
\begin{equation}
E_{pb}^{L=1}\approx 2 E^\prime_{ann} \approx (20\sim 60)
\mbox{MeV}
\end{equation}
The presence of the orbital excitation in $\Theta$ pentaquark
contributes an additional energy $E_L \approx 240$ MeV to its
mass. However the centrifugal barrier from the orbital excitation
reduces the Pauli blocking energy from $2E_{pb}^{L=0}\approx 430$
MeV to $2E_{pb}^{L=1} \approx (20\sim 60)$ MeV. This effect and
the annihilation effect $-4 E^\prime_{ann}$ work together to make
$\Theta^+$ pentaquark a low lying baryon.

The singlet pentaquark mass reads
\begin{equation}\label{11}
M_1 = 1662-2E_{ann}-2E^\prime_{ann}= (1402\sim 1542) \mbox{MeV}
\end{equation}
Clearly this $P=-$ pentaquark singlet state is very probably
low-lying in the framework of diquark model. Possible decay
channels were suggested for future experimental searches in
\cite{zhang}.

Putting everything together, we get a rough estimate of the P=-
dibaryon mass:
\begin{equation}
M_{\Lambda_6}= (2270\sim 2310) \mbox{MeV}
\end{equation}
This P=- isoscalar dibaryon state is probably $(40\sim 80)$ MeV
above $\Lambda\Lambda$, $\Xi N$ threshold. So it's unstable
against P-wave $\Lambda\Lambda$ and $\Xi N$ strong decays. But
it's possibly stable against $\Xi N \pi$ or $\Sigma \Lambda\pi$
S-wave strong decays. Its width is expected to be not very broad.
This state could be searched at RHIC.

The author thanks F. E. Close and Q. Zhao for helpful
communications. This project was supported by the National Natural
Science Foundation of China under Grant 10375003, Ministry of
Education of China, FANEDD and SRF for ROCS, SEM.


\end{document}